\documentclass[aps,preprint,showpacs]{revtex4}
\usepackage{epsfig,braket}

\begin{document}

\title{Design of realistic switches for
coupling superconducting solid-state qubits}

\author{Markus J.\ Storcz}
\email{storcz@theorie.physik.uni-muenchen.de} \author{Frank K.\
Wilhelm}

\affiliation{Sektion Physik and CeNS,
Ludwig-Maximilians-Universit\"at,  Theresienstr.\ 37, 80333 M\"unchen,
Germany}

% \date{DO NOT DISTRIBUTE THIS DRAFT OF \today}

\begin{abstract}
Superconducting flux qubits are a promising candidate for solid-state quantum
computation. One of the reasons is that implementing a controlled coupling
between the qubits appears to be relatively easy, if one uses 
tunable Josephson junctions. We evaluate possible coupling strengths and
show, how much extra decoherence is induced by the subgap conductance of 
a tunable junction. In the light of these results, we evaluate several options
of using intrinsically shunted junctions and show that based on available
technology, Josephson field effect transistors and high-T$_c$ junctions
used as $\pi$-shifters would be a good option, whereas the use of magnetic
junctions as $\pi$-shifters severely limits quantum coherence. 
\end{abstract}


\maketitle

%\section{Introduction}

Quantum computation promises qualitative improvement of computational power 
as compared to today's classical computers. An important candidate
for the implementation of a scalable quantum computer are superconducting
qubits \cite{Makhlin,orlando}. After experimental demonstration of
basic features e.g.\ in flux qubits 
\cite{Caspar,Chiorescu}, the improvement of the
properties of such setups involves 
engineering of couplings and decoherence, see e.g.\ 
\cite{EPJB}. 

\begin{figure}[ht]
\begin{center}
 \includegraphics*[width=16cm]{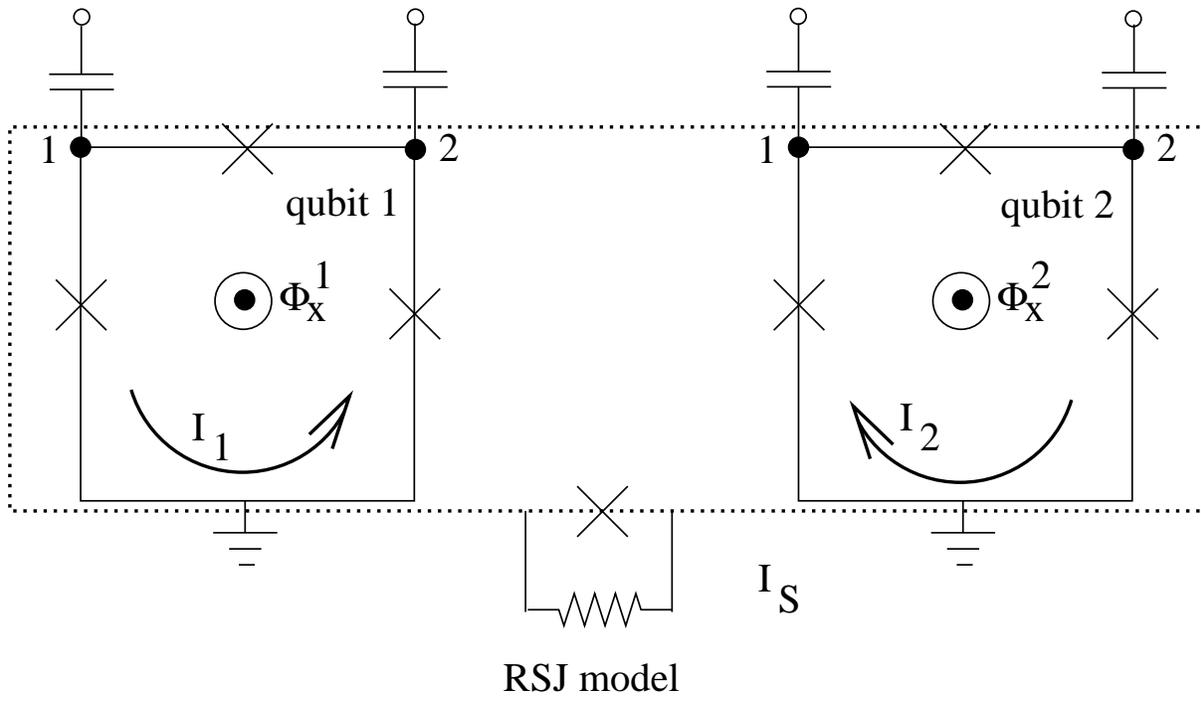}
\end{center}
\caption{The flux transformer inductively couples two flux qubits \cite{orlando}.
It can be switched, e.g. by a DC-SQUID or by a tunable shunted Josephson junction.} \label{transformer}
\end{figure}

To perform universal quantum computation 
with a system of coupled qubits it is very much desirable to be able
to switch the couplings (although there are in principle
workarounds \cite{untunable}). It has already been described that
for flux qubits, this can be achieved by using a superconducting
flux transformer interrupted by a tunable Josephson junction 
\cite{orlando},
i.e.\ a superconducting switch, as shown in Fig.\ \ref{transformer}.
The primary and most straightforward proposal for the implementation of 
this switch is to use an unshunted DC-SQUID based on tunnel junctions utilizing
the same technology as for the qubit junctions. Although this holds the promise
of inducing very little extra decoherence, 
it suffers from two practical restrictions: i) the SQUID loop
has to be biased by exactly half a flux
quantum in the off-state and ii) the external control parameter is a 
magnetic flux, which introduces
the possibility of flux cross-talk between the qubits and the switch. The 
combination of i) and ii) implies that even small flux cross-talk will 
severely perturb
the off-state of the switch. 

This can be avoided by using different switches: A voltage-controlled
device such as a Josephson Field Effect Transistor (JoFET) \cite{richter} or 
an SNS-Transistor completely avoids
the cross-talk problem. As an intermediate step \cite{Blatter}, one can improve the
SQUID by using a large 
$\pi$-Junction, in order to fix the off-state at zero field. 
Such $\pi$-junctions can
be found in high-T$_c$ systems \cite{hightc} 
or in systems with a magnetic barrier \cite{sfs}. All these junctions 
are damped by a large subgap conductance because
they contain a large number of low-energy quasiparticles. 

In this letter, we quantitatively evaluate the coupling strength between
two qubits coupled by a switchable flux transformer. We evaluate the
strength of the decoherence induced by the subgap current modeled
in terms of the RSJ model. Based on this result, we assess available 
technologies for the implementation of the switch.
% \begin{figure}[ht]
% \begin{center}
% \includegraphics[width=7.5cm]{transformer1.eps}
% \end{center}
% \caption{The flux transformer inductively couples two flux qubits \cite{orlando}.
% It can be switched, i.e. by a DC-SQUID or by a tunable shunted Josephson junction.} \label{transformer}
% \end{figure}

We start by calculating the strength $K$ of the coupling between the
two qubits without a switch and then show how it is modified by the 
presence of the switch.
%\subsection{No switch}
From Fig. \ref{transformer} and the law of magnetic induction we find the 
following equations for the flux
through qubit 1 and 2 induced by currents in the qubits and the flux 
transformer
\begin{eqnarray}\label{flux_current_relm1}
\delta\left( \begin{array}{c}
\Phi_S \\
\Phi_1 \\
\Phi_2
\end{array} \right)
& = &
\left( \begin{array}{ccc}
M_{TT} & M_{TQ} & M_{TQ} \\
M_{TQ} & M_{QQ} & 0 \\
M_{TQ} & 0 & M_{QQ}
\end{array} \right)
\left( \begin{array}{c}
I_S \\
I_1 \\
I_2
\end{array} \right)\textrm{.}
\end{eqnarray}
where $M_{QQ}$ is the self-inductance of the qubits (assumed to be identical),
$M_{TQ}$ is the mutual inductance between
the transformer and the qubits and the mutual inductance
between the qubits is assumed to be negligible.
The fluxes $\delta\Phi$ in Eq.\ (\ref{flux_current_relm1}) are the screening fluxes in the
transformer and the two qubits, i.e.\ the deviations from the externally
applied values.
Henceforth,
we abbreviate Eq.\ (\ref{flux_current_relm1}) as
$\delta \vec \Phi  =  \mathbf{M} \vec I$.
These formulas are general and can be applied for any flux through
the transformer loop. It is most desirable to couple zero net
flux through the device, which can be achieved by 
using a gradiometer configuration
\cite{tinkham}.
For this gradiometer case, we get
$I_S = - (M_{TQ}/M_{TT}) (I_1+I_2)$,
which we might insert into (\ref{flux_current_relm1}) and find for the
inductive energy
\begin{equation}
E_{\rm ind} 
	=  \left( M_{QQ} - \frac{M_{TQ}^2}{M_{TT}} \right) ( I_1^2 + I_2^2)
	- 2 \frac{M_{TQ}^2}{M_{TT}}I_1 I_2 \textrm{.}
\end{equation}
The terms resulting from the off-diagonal elements of (\ref{flux_current_relm1}) can
directly be identified as the interqubit coupling strength $K= -2 (M_{TQ}^2/M_{TT}) I_1 I_2$ which enters the $\hat \sigma_z\otimes \hat \sigma_z$ Ising-coupling
described in Refs.\ \cite{orlando,PRA}. 
 Note, that the dynamics of the qubit flux
is dominated by the Josephson energies \cite{orlando}, to which 
the diagonal term is only a minor correction. 

%\subsection{Tunable Josephson junction as a switch}
We now introduce the tunable Josephson junction into the loop. 
Using fluxoid quantization,
we rewrite the Josephson relation \cite{tinkham}
$I_S = I_c \sin \left( -2\pi (\Phi_S/ \Phi_0) \right)$ and insert it into
eq.\ (\ref{flux_current_relm1}). The resulting nonlinear equation can 
be solved in the following cases: i) 
If $\vert I_S/I_c \vert \ll 1$ (``on'' state of the switch) 
we find $K  =  - 2 (M_{TQ}^2/M_{TT}^\star )I_1 I_2$ 
with
$M_{TT}^\star := M_{TT}+(\Phi_0/2\pi I_c)= M_{TT}+L_{\rm kin}(0)\textrm{.}$
This can be understood as an effective increase of the self-inductance
of the loop by the kinetic inductance of the Josephson junction at
zero bias. ii) In the case 
$\vert I_S/I_c \vert \approx 1$, ``off'' state, the circulating current
is close to the critical current of the switch, hence the phase drop is
$\pm \pi/2$ and we  find an analogous form
$K  =  - 2 (M_{TQ}^2/M_{TT}') I_1I_2$ with 
$M_{TT}' = M_{TT}+(\Phi_0/4 \vert I_c \vert)$, i.e.\ at
low $I_c$ the coupling can be arbitrarily weak due to the enormous kinetic 
inductance of the junction close to the critical current. 

%\subsubsection{RSJ-model}
We now turn to the discussion of the decoherence induced by
the subgap conductance of the tunable junction. The decoherence occurs due
to the flux noise generated through the current noise from the
quasiparticle shunt. Hence, both qubits experience the same level
of noise. The decoherence of such a setup has been extensively studied
in Ref.\ \cite{PRA} as a function of the environment parameters. In this letter,
we evaluate these environment parameters for our specific setup.

\begin{figure}[ht]
\begin{center}
\includegraphics*[width=8cm]{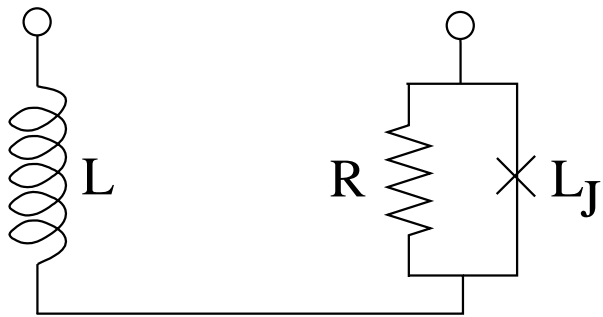}
\end{center}
\caption{Equivalent circuit diagram of the flux transformer circuit. The JoFET is modeled by
a resistively shunted Josephson junction.} \label{transformer_2}
\end{figure}

We model the junction 
by the RSJ-model for a sound quantitative estimate
of the time scales even though the physics of the subgap conductance is
usually by far more subtle than that.
We evaluate the 
fluctuations of the current between two points of the flux transformer loop sketched
in figure \ref{transformer}. $L$ is the geometric inductance of the loop, $L_J$
is the Josephson inductance characterizing the Josephson contact and $R$ is the shunt
resistance.
% \begin{figure}[ht]
% \begin{center}
% \includegraphics{transformer2.eps}
% \end{center}
% \caption{Equivalent circuit diagram of the flux transformer circuit. The JoFET is modeled by
% a resistively shunted Josephson junction.} \label{transformer_2}
% \end{figure}
The correlation is given by the fluctuation-dissipation theorem
$\braket{\delta I \delta I}_\omega = \mbox{coth} ( \beta \hbar \omega/2 ) \hbar \omega
{\rm Re} Y(\omega) \textrm{,}$
where $Y(\omega)$ is the admittance of the effective circuit depicted 
in Fig. \ref{transformer_2}. 
Following the lines of Ref.\ \cite{EPJB}, this 
translates into a spectral function of the energy fluctuations of the
qubit of the shape $\langle \delta\epsilon(t)\delta\epsilon(0)\rangle_\omega
=J(\omega)\coth(\hbar\omega/2k_BT)$ with $J(\omega)=\alpha \omega^2/(\omega^2+\omega_c^2)$
with the important result that the dimensionless dissipation parameter here reads
\begin{equation} \label{alpha}
\alpha = \frac{4 I_{\rm circ}^2M_{TQ}^2L_J^2}{hR(L+L_J)^2}
\end{equation}
and a cutoff $\omega_c=R(L+L_J)/LL_J$. Here, $L_J=\Phi_0/2\pi I_c$ is
the kinetic inductance of the junction. From (\ref{alpha}) we receive in the limit
$L \gg L_J$ the expression $\alpha \propto 1/RI_c^2$ and for
$L \approx L_J$, $L \ll L_J$ it follows that $\alpha \propto 1/R$.
From the results of Ref.\ \cite{PRA}, we can conclude that 
 $\alpha \approx 10^{-6}$ 
poses an
upper bound for gate
operations to be compatible
with quantum error correction.
In the following sections we will evaluate $\alpha$ for different types 
of junctions in the switch, 
a JoFET,
an SFS junction and a high-$T_c$ junction by inserting typical parameters. 
We use the normal resistance $R_N$ to estimate the shunt resistance in
the RSJ model.
Here, it is
important to note that the parameters $I_c$ and $R_N$
of the junction determine the suitability of the device as a (low-noise) 
switch, which are given by a combination of material and geometry
properties. In the following we exemplify the calculation of the
dissipative effects with several experimental parameter sets.
% \begin{figure}[t]
% \begin{center}
% \includegraphics*[width=8cm]{jofet_switch.eps}
% \end{center}
% \caption{The dimensionless dissipation parameter $\alpha$ as a function of
% the electron density in the 2DEG for a JoFET.
% The inset shows a linear plot of the region with the largest $\alpha$.} \label{jofetswitch}
% \end{figure}

% \begin{figure}[t]
% \begin{center}
% \includegraphics*[width=8cm]{icrnalpha.eps}
% \end{center}
% \caption{Log-log plot of normal state resistance versus the
% critical current of the junction. Here $R_N$ is taken as an estimate
% for the shunt resistance of the junction. The solid line denotes $\alpha=10^{-6}$ and
% the two dotted lines are for $\alpha=10^{-4}$ (lower line) and $\alpha=10^{-8}$ (upper
% line). Parameters for the SIFS-junction are $I_c \approx 8.5 \cdot 10^{-5}$ A and
% $R_N \approx 250$ m$\Omega$ \cite{kontos}.} \label{icrnalpha}
% \end{figure}

For present day qubit technology \cite{Bhuwan} we can assume
$L \approx 1$ nH, $I_{\rm circ} \approx 100$ nA
$M_{TQ} \approx
100$ pH. In the following, we estimate $\alpha$ for a number of junction 
realizations, adjusting the junction area for sufficient critical current.

%\subsubsection{Josephson field-effect transistor (JoFET)} \label{secjofet}
A Josephson field-effect transistor (JoFET) can be understood as an 
SNS junction where the role of the normal metal is played by a doped
semiconductor.
By applying a gate voltage, it is possible to tune the electron density of the 
semiconductor. 

The critical current of such a junction containing $N_{\rm ch}$ channels 
can be found using 
the formula of Kulik and Omel'yanchuk
$I_c = (\pi \Delta)/(R_N e)$ \cite{tinkham,kulik_omelyanchuk}. $R_N=h/(2e^2N_{\rm ch})$
is the point-contact resistance. In a JoFET, the back gate essentially
controls $N_{\rm ch}$.
The typical 
normal resistance is around $R_N \approx 10$ $\Omega$.
For a JoFET the critical current of the Josephson junction is $I_c \approx 30$ $\mu$A and the
Josephson inductance is $L_J \approx 11$ pH \cite{richter}.

Inserting the above estimates we get
$\alpha \approx 7 \cdot 10^{-6}\textrm{.}$
This means that the dissipative effects are weak and a JoFET should
be a reasonable switch that poses no new constraints.
Besides the obvious technological challenge
\cite{richter}, one 
drawback of JoFETs is that due to wide junctions with dimensions
of around $w=500$ nm they are likely to trap vortices, which
can cause 1/f noise by hopping between different pinning sites.
However, this can be reduced by pinning e.g.\ by perforating the junction.

\begin{figure}[t]
\begin{center}
\includegraphics*[width=16cm]{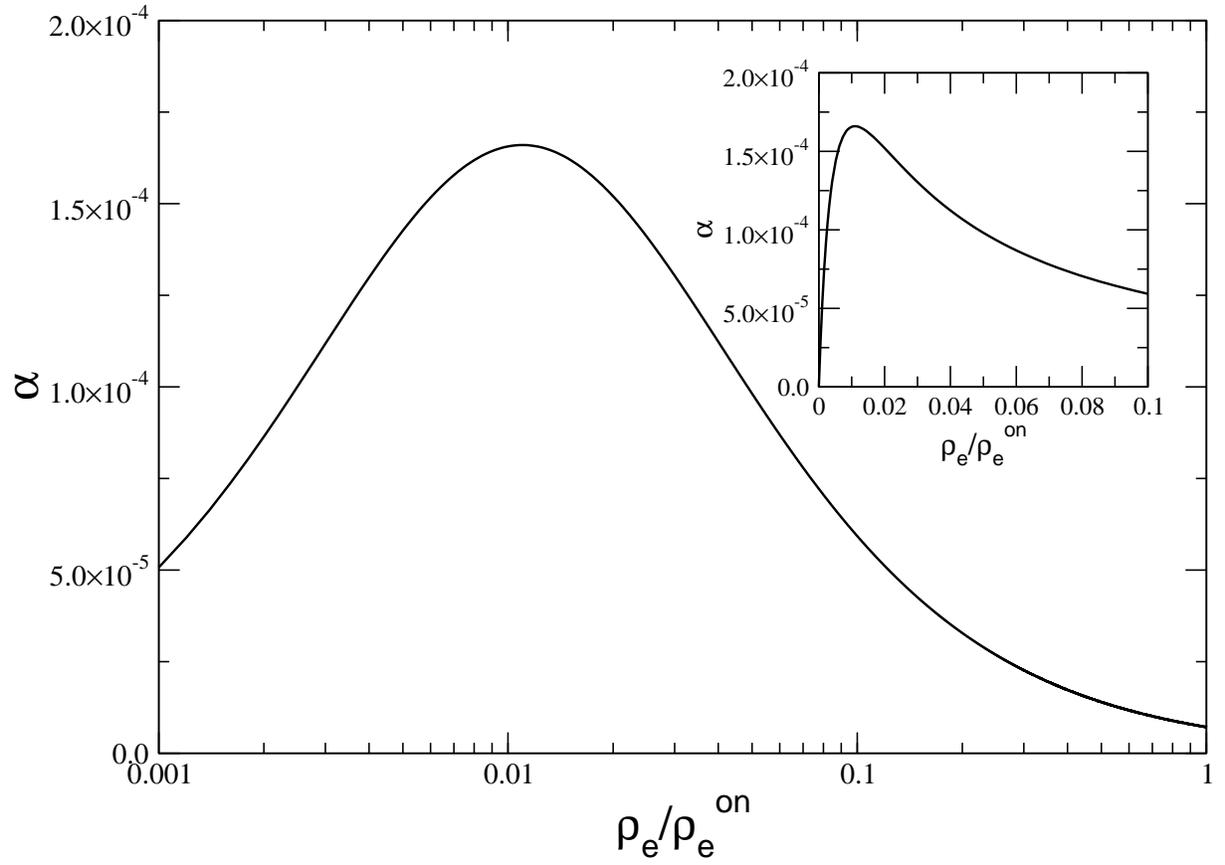}
\end{center}
\caption{The dimensionless dissipation parameter $\alpha$ as a function of
the electron density in the 2DEG for a JoFET.
The inset shows a linear plot of the region with the largest $\alpha$.} \label{jofetswitch}
\end{figure}

If we go away from the ``on'' state with the JoFET, we reduce both
$I_c$ and $G_N$ linearily by depleting the density of states.
Fig.\ \ref{jofetswitch} shows that we find that the dissipative effects are strongest during the switching
process when $L_J(\rho_e/\rho_e^{on}) \approx L_{J,0}$, and {\em not} in the ``on'' state of
the switch. In the ``off'' state of the switch (for $\rho_e(0)\rightarrow 0$) also $\alpha$ goes to zero. If the switch is tuned
from the ``off'' state to the ``on'' state, $\alpha$ reaches a local maximum and then decreases
again. This makes the JoFET a very attractive switch: It induces an 
acceptably low level decoherence in the ``on'' state and can be made
completely silent in the ``off'' state.

\begin{figure}[t]
\begin{center}
\includegraphics*[width=16cm]{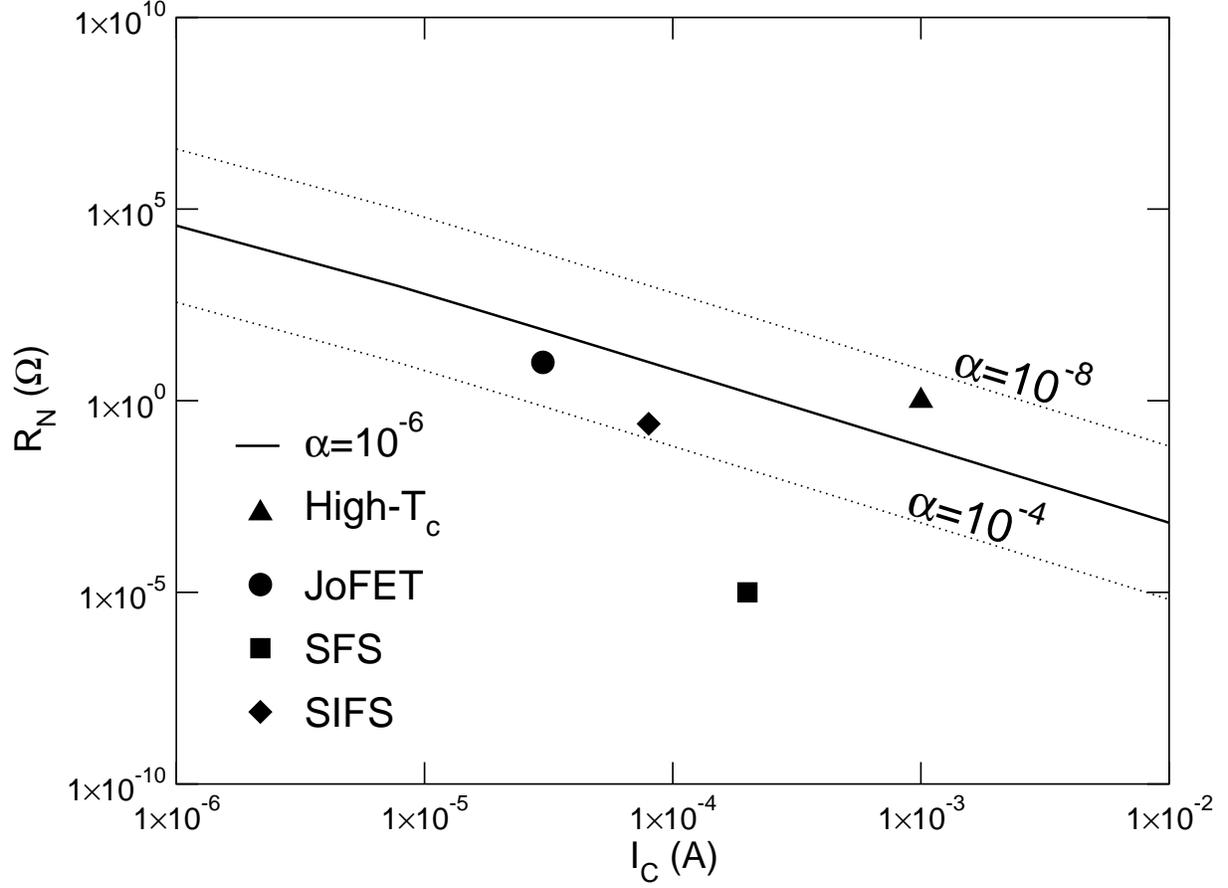}
\end{center}
\caption{Log-log plot of normal state resistance versus the
critical current of the junction. Here $R_N$ is taken as an estimate
for the shunt resistance of the junction. The solid line denotes $\alpha=10^{-6}$ and
the two dotted lines are for $\alpha=10^{-4}$ (lower line) and $\alpha=10^{-8}$ (upper
line). Parameters for the SIFS-junction are $I_c \approx 8.5 \cdot 10^{-5}$ A and
$R_N \approx 250$ m$\Omega$ \cite{kontos}.} \label{icrnalpha}
\end{figure}

%\subsubsection{SFS junction}
An SFS junction in the $\pi$-state is based on a metallic material, thus
the estimate of the shunt resistance in the
RSJ model yields a much smaller result than in the case of the JoFET, $R \approx 10^{-5}$ $\Omega$
\cite{sfs}. 
The critical current of the SFS
junction is $I_c \approx 0.2$ mA. Thus, leaving the transformer
properties unchanged, we find $L_J \approx 1.7$ pH.
Using these 
estimates the strength of the dissipative effects is of the order of
$\alpha \approx 0.16$. This makes such a device unsuitable at the
present level of technology, however, it appears that SIFS junctions \cite{kontos} are by far closer to the desired values, see Fig.\ \ref{icrnalpha}.

%\subsubsection{High-$T_c$ junction}
High-$T_c$ junctions can be realized in different ways.
Here, we take from Ref.\ \cite{hightc} parameters for a typical noble
metal (Au)-bridge
junction with a film thickness of about $w \approx 100$ nm.
The product $I_c R_N \approx 1$ mV and $\rho_N=8.3$ $\Omega$ nm.
We assume that in principle $I_c$ for the
$\pi$-state and the 0-state are the same.
For a contact area of around $30$x$30$ nm$^2$, $I_c \approx 1 $ mA and $R_N \approx 
1$ ${\rm \Omega}$. Now the strength of the dissipative effects
is easily evaluated to be $\alpha \approx 6.5 \cdot 10^{-8}$, which
is way better than SFS $\pi$-junctions and even better than the JoFET.

We estimated the strength of the dissipative effects that will occur due to the switch
for several possible switches. These results are summarized in figure \ref{icrnalpha}
for typical parameters of the analyzed systems. We find that the noise properties
of a JoFET and $\pi$-shifters based on High-T$_c$-materials introduce no important
noise source. On the other hand, the parameters found from $\pi$-shifters based on magnetic
materials are much less encouraging.

We would like to thank T.P.\ Orlando, P.~Baars and A.~Marx for useful discussions. Work
supported by ARO through Contract-No. P-43385-PH-QC.

\newpage

\end{document}